\documentclass[twocolumn,showpacs,preprintnumbers,prl,floatfix]{revtex4-1}
\usepackage{amssymb,amsmath,amsfonts}
\usepackage{graphics,dcolumn,bm,fleqn,epic,eepic,float}
\usepackage{graphicx} 
\usepackage{epstopdf} 
\usepackage{bm} 
\usepackage{subfigure}
\usepackage{color} 

\begin{document}
\title{How women organize social networks different from men}

\author{Michael Szell$^{1}$, Stefan Thurner$^{1,2,3}$}

\affiliation{%
  $^1$~Section for Science of Complex Systems, Medical University of
  Vienna, Spitalgasse 23, 1090 Vienna, Austria}
\affiliation{%
  $^2$~Santa Fe Institute, Santa Fe, NM 87501, USA}
\affiliation{%
  $^3$~IIASA, Schlossplatz 1, 2361 Laxenburg, Austria}

\begin{abstract} 
Superpositions of social networks, such as communication, friendship, or trade networks, are called multiplex networks, forming the structural backbone of human societies. Novel datasets now allow quantification and exploration of multiplex networks. Here we study gender-specific differences of a multiplex network from a complete behavioral dataset of an online-game society of about 300,000 players. On the individual level females perform better economically and are less risk-taking than males. Males reciprocate friendship requests from females faster than vice versa and hesitate to reciprocate hostile actions of females. On the network level females have more communication partners, who are less connected than partners of males. We find a strong homophily effect for females and higher clustering coefficients of females in trade and attack networks. Cooperative links between males are under-represented, reflecting competition for resources among males. These results confirm quantitatively that females and males manage their social networks in substantially different ways.
\end{abstract}


\maketitle

\begin{figure}[!t]
\begin{center}
\includegraphics[width=0.45\textwidth]{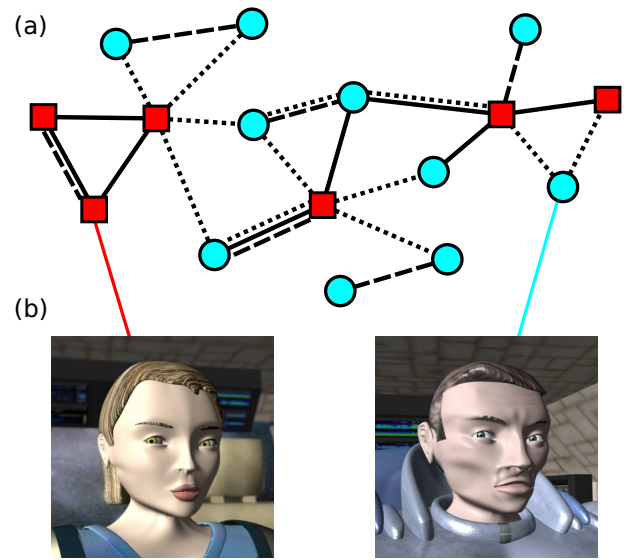}
\end{center}
\caption{
{\bf Multiplex networks consist of a set of nodes connected by different types of links.} (a) In the MMOG dataset, six types of social links co-exist between players, representing their friendship 
or enmity status,  exchange of messages, trading activity,  aggressive acts (attacks), 
and their placing of head-money (bounties) on others as a means of punishment. 
(b) Each player of the MMOG {\em Pardus} chooses a male (circles) or female (squares)  
gender when joining the game. 
\label{fig:multiplexitygender} 
}
\end{figure}

Essential to understanding the behavior of humans within their socio-economical environment is the observation 
that they simultaneously play different roles in various interconnected social networks, such as friendship networks, 
communication networks, family,  or business networks. The superposition of several networks on the same 
set of nodes (individuals) is called a {\em multiplex network} (MPN), Fig. \ref{fig:multiplexitygender}. 
In social networks, interactions between individuals (such as communication, trading, aggression, revenge) 
can be represented as links. Since these interactions can be unilateral and do not necessarily have to be reciprocated, 
in general the MPN consists of a set of directed, weighted and time-varying subnetworks. The MPN is a highly 
non-trivial, dynamical object which allows to quantify societies on a systemic level. In particular the MPN captures 
the co-evolving nature of societies, where on one hand  actions of individuals shape and define the topological 
structure of the MPN, and on the other hand, where the topology of the MPN constrains and shapes the possible actions. 
The position of an individual within the MPN has an  impact on her/his  possibilities for actions in the near future 
and her/his {\em state} of being. The state of an individual characterizes its present situation such as its wealth, 
achievements, skills, number of friends, etc. The notion of the MPN conveniently allows to study the interdependencies 
of different social relations and in particular how they mutually influence each other through  {\em network-network 
interactions} \cite{szell2010mol}. 

The study of small-scale MPNs has an established tradition in the social sciences \cite{wasserman1994sna, 
mcpherson2001bfh, entwisle2007ncv, padgett1993rar}, for large societies however, in general the MPN can 
not be observed directly due to immense data requirements. Nevertheless there are recent considerable achievements 
in understanding several massive social networks on a quantitative basis, such as the cell phone communication network 
\cite{onnela2007sts, onnela2007als, lambiotte2008gdm}, features of the world-trade network \cite{hidalgo2009bbe, 
hidalgo2007psc, klimek2011ecc}, email networks \cite{newman2002ens}, the network of financial debt \cite{boss2004nti} 
and the network of financial flows \cite{kyriakopoulos2009nea}. The integration of various time-varying networks on 
an entire society has so-far been beyond the scope of any realistic study.

Online media have revolutionized the possibilities of measuring human behavior on societal levels and allow for the possibility of 
transforming the social sciences into a quantitative, experimental science \cite{lazer2009css, lewis2008ttt, thurner2012egc}. 
Interactive online games provide a particularly powerful tool to quantitatively study human socio-economic multiplex data 
\cite{bainbridge2007srp,castronova2005swb}.  
These games are usually played by thousands, sometimes even millions and offer `alternative lives' where
players engage in different types of social interactions, such as establishing friendships and economic 
relations,  forming of groups, alliances, parties, or performing aggressive acts like fighting or waging war 
\cite{szell2010msd}. Practically all actions of all players can be recorded at a time resolution of a second.
Massive Multiplayer Online Games (MMOGs) open a previously unimaginable potential for data-driven, quantitative 
socio-economic understanding of  human societies \cite{castronova2006orv,johnson2009hgf,szell2010msd,szell2010mol},
with three  obvious  advantages over traditional methods:  
 (i) Complete information on the {\em topology of the MPNs}: 
 All topological properties of the  MPN are available. 
 The inter-dependence of the different networks in the MPN reveals the organization of the system at different levels \cite{szell2010mol}.
 (ii) Complete information about the {\em co-evolution of the MPN} with the behavior, actions and performance of individual players. 
 Players can be grouped into different classes such as their gender so that the MPN can be studied gender-specifically.  
 (iii) {\em Reduced measurement bias}: players are not constantly aware of their actions being recorded -- the experimental setup does not influence the 
 behavior which is often a  problem in behavioral experiments \cite{henrich2005emc}.  

Humans actively influence and re-structure their local MPN. In fact a large fraction of human activity is devoted to re-arranging, managing and 
maintaining various social networks at the same time. 
In this paper we focus on gender-specific differences in managing and maintaining local MPN environments and how this correlates with performance. 
To identify differences in networking behavior of male and female players, we analyze complete and coherent MPN data from a MMOG 
society of about 300,000 players \cite{castronova2005swb}.

There exist a few quantitative, data-driven studies on gender-specific behavior with the focus on the use of communication media, 
including a study on telephone usage in 317 French homes  \cite{smoreda2000gud}, 
a large-scale Belgian mobile phone call and test message dataset \cite{stoica2010agc}, and a gender-specific 
content analysis of instant messages \cite{fox2007mmd, baron2004syo}. Gender differences in online gaming is relatively unexplored, 
with the exception of recent work on gender roles \cite{williams2009lgg}, the relation of gender 
and age group \cite{griffiths2004ocg}, and gender swapping \cite{griffiths2003bsc, bruckman1993gsi, danet1996tmg, hussain2008gss}, 
i.e. the phenomenon of playing a gender different from the biological sex.

Pardus (\href{http://www.pardus.at}{http://www.pardus.at}) is a browser-based MMOG in a science-fiction setting, 
played since September 2004. MMOGs are characterized by a substantial number of users playing together online
in the same virtual environment \cite{bartle2004dvw,castronova2005swb}. In Pardus every player owns a single, 
individual account which is associated with a game \emph{character}. A character owns a spacecraft with a certain 
cargo capacity, and can roam the virtual universe, trade commodities, socialize, and do much more, usually with 
the motivation to gain wealth, social standing, respect, and fame in the virtual world  
(\href{http://www.pardus.at/index.php?section=about}{http://www.pardus.at/index.php?section=about}). 
One feature of Pardus is a trading platform which provides the possibility for the production and distribution of virtual 
goods and services. The `society' of Pardus is mainly driven by social factors such as friendship, cooperation or 
competition. Its gameplay is based on socializing and role-playing, with interaction between players, and with interactions 
between  players and non-player characters \cite{castronova2005swb}. 
It is important to mention that the various social networks which can be extracted from the Pardus game indeed form a 
MPN, in the sense that these networks are tightly related and mutually influence each other. This was systematically 
explored and quantified in \cite{szell2010mol}.
For  more details on Pardus, see \cite{szell2010msd}.

When signing up for the game the player choses to be a male or female character. The gender of a character can not 
be changed afterwards. The consequence of the virtual gender is that a male or female avatar is displayed during 
communication in the game, see Fig. \ref{fig:multiplexitygender} (b). We have no information about the biological sex of players.  
The possibility of freely choosing a virtual sex may lead some players to experiment with gender roles \cite{danet1996tmg, 
bruckman1993gsi, bartle2004dvw, hussain2008gss}. Selecting a gender different from the biological is called 
\emph{gender swapping} which is common in online environments \cite{grosman2010esg}. A survey on 8,694 players 
in the  \emph{Everquest} MMOG found that 15.5\% are gender-swapped, 17\% of the males and 10\% of the  females 
\cite{griffiths2003bsc}. Similar values are reported for the virtual environment  \emph{Second life} (10\% swapping of all, 
16\% of males, just 2.7\% of females) \cite{de2006sls} or the \emph{World of Warcraft} MMOG (23\% of males, 3\% of 
females swapping for the `most enjoyable' character) \cite{yee2007pee}.
In the Pardus game the data acquisition process aggregates both possible gender-specific behavioral effects resulting from 
the underlying biological sex of the players on one hand, and from their chosen online gender -- regardless of sex -- on the other hand.

Players can anonymously mark others as \emph{friends} or \emph{enemies}. Private Messages (PM) are the prevalent 
form of communication in the game. It is comparable to email in the real world. A PM is seen by sender and receiver only. 
For more details see \cite{szell2010msd}. Communication events, friend and enemy markings, trades, attack and  bounty 
placements are recorded as networks: A link of type `communication', `friend' etc. is placed from player $i$ to $j$ if  $i$ 
has communicated with $j$, or has marked $j$ as friend/enemy, etc. Friend/enemy markings exist until they are removed 
by players, while PM, trade, attack, and bounty networks are recorded by aggregating all actions that happened before to 
a given day. Friend and enemy networks are unweighted, for all others the weights  correspond to the cumulative number 
of PMs sent, trades made, etc. The actions of communication, trading, and friend marking we consider to be positive (friendly) 
actions, whereas attack, enemy marking and bounty collection are negative (unfriendly). For more information on the MPN 
data, see SI. The legal department of the Medical University of Vienna has attested the innocuousness of the used anonymized 
data.

\section{Results}

\begin{table}[!t]
\caption{
{\bf Quantifiable achievements of players in the game.} These include spent APs, earned credits, experience points earned, players killed, deaths,
bounties collected. 
For each player we calculate a respective daily value for these quantities, 
and average over them. The respective units are used as in the game (arbitrary units).
The last column denotes whether the null hypothesis $H_0$ of equal means is rejected.
Females collect significantly more money  (4-sigma event) and experience less  deaths (2 sigmas) than males. 
\label{tab:achievements}
}
\begin{center}
\begin{tabular}{l c c c}
				& 	Female	&	Male 	 & $H_0$ rejected		\\ \hline
Activity (APs)		&	2714 	&	2607   & no			\\
Wealth (Credits)	&	14619&	11073	     &    yes ****\\ 
Experience		&	270 	&	250	& no		\\
Kills				&	0.0055	&	0.0063	& no	\\
Deaths			&	 0.026	&	0.030  & yes * \\
Bounties 		 	&	0.0024	&	0.0025  & no		\\ \hline
\end{tabular}
\\ * p-value $<$ 0.05, **** p-value $<$ 0.0001
\end{center}
\end{table}

{\bf Females are less risk-taking but wealthier.}
By economic actions such as trade, players earn virtual money in the currency of \emph{credits}. 
Destructive, and aggressive actions such as attacking others, may result in the destruction of those 
players (kills),  the destruction of oneself (death), or the collection of a bounty if a bounty was placed 
on a destroyed player's head. Players spend \emph{action points} (AP) to perform all of these actions, 
which provides a measure for their overall activity. Players have the possibility to earn 
\emph{experience points} through various actions or jobs, such as destroying `space monsters'.

In Table~\ref{tab:achievements} we report the number of APs spent, credits and experience points earned, 
players killed, deaths, and bounties collected per day, averaged over all players. 
To compare values from males with values from females, we adopt a standard two-sample t-test, testing for the null 
hypothesis $H_0$ that the two distributions have equal means, against the alternative hypothesis $H_1$ that the 
means are not equal. The test is two-tailed, making no assumptions whether values from females are supposed to 
be larger than from males or vice versa, see Methods. It is immediately clear that females accumulate significantly more 
wealth ($H_0$ rejected at 4 sigma level), i.e. credits, than males. At the same time male players experience significantly 
more deaths (rejection at 2 sigmas), due to more risk-taking and/or aggressive behavior. This points at a much larger 
engagement of females in economic, rather  than destructive activities. Concerning overall activity, experience points, 
kills and collected bounties, female and male players perform comparably  ($H_0$ cannot be rejected).

\begin{table*}[!t]
\caption{{\bf Gender-specific number of relational actions and links.}
Upper panel:
Out-weights (actions initiated or messages sent) and in-weights (actions/markings/messages received). 
For each player we calculate a daily value for the number of actions performed and received and average them.
Male values are presented as in Table I. 
For values with stars the null hypothesis $H_0$ of equal means is rejected.
Females mark significantly more friends ($H_0$ rejected at 2 sigmas) and are marked themselves much more as friend 
(5 sigmas). They send and receive significantly more PMs (3 and 5 sigmas, respectively), and perform and receive significantly
more trades (4 and 5 sigmas). Females perform less attacks than males ($2$ standard deviations). 
Lower panel:
Homophily and heterophily: Over- and under-represenation, as measured by Z-scores (See SI), of directed link types 
of the six networks at day 856. Each real network is compared to 1,000 surrogate networks where the gender
of nodes is randomly reshuffled. The most significant over-represented link types are female-to-female trades and communication 
(PMs), followed by male-to-female trades and PMs. Female-to-male trades and PMs are over-represented, however not as drastically. 
Highly under-represented are male-to-male PMs and trades. Negative links show neither substantive over- or under-representation.
\label{tab:weights} }
\begin{center}
\begin{tabular}{l ccc |ccc}
  & \multicolumn{3}{c|}{Positive links} & \multicolumn{3}{c}{Negative links}\\ \hline

  & Friends &  PMs &  Trades &  Enemies & Attacks &  Bounties  \\
  Action weights &&&&&&\\  
\hline   
Initiated M 	& 0.026  & 0.60 &  0.100  & 0.029 & 0.20 & 0.009 \\
Initiated F 	& 0.028*   &  0.74**  &  0.124***  &  0.027  &  0.16*  & 0.009  \\[0.15cm]
Received M & 0.022 & 0.45 & 0.066 & 0.036 & 0.116 & 0.0084\\
Received F & 0.027**** & 0.61**** & 0.086**** & 0.032 & 0.109 & 0.0074 \\[0.25cm]

Homophily & & &&&&\\
\hline
                   FF &  0.6 &  3.9 &  4.2 &  0.6 & -0.4 & -0.6 \\
		  MM & 0.4 & -2.3 & -2.4 & -0.6 &  1.0 & -0.2 \\
                   FM & -0.4 &  1.6 &  1.4 & -0.3 & -1.7 &  0.4 \\
                   MF & -0.7 &  2.7 &  2.7 &  0.7 &  0.7 & -0.2 \\ \hline
\end{tabular}
\end{center}
    * p-value $<$ 0.05, ** p-value $<$ 0.01, *** p-value $<$ 0.001, **** p-value $<$ 0.0001
\end{table*}

{\bf Females attract positive behavior.} We now focus  on  action weights, i.e. how many actions of each type 
are initiated (out-weight) or received (in-weight) by the average female or male player. Table~\ref{tab:weights}, 
upper panel, shows the out- and in-weights for the different action types. 
Highly significant differences for all positive action types are observed, where females are much more active in 
performing actions: marking friends (null hypothesis $H_0$ of equal means rejected at 2 sigmas), writing private 
messages (3 sigmas), or initiating a trade (4 sigmas). At even higher significance levels (rejection of $H_0$ at 5 
sigmas) females receive positive actions with respect to males: being marked as friend, receiving a private message, 
and being on the receiving end of a trade than their male co-players. For all negative action types males are both 
more active in initiating and receiving than females, however to a less substantial degree.

{\bf Females show homophily, males are heterophiles.} 
Homophily is the tendency of individuals to associate and link with similar others \cite{mcpherson2001bfh}.
In our case we test for linkings between individuals of the same gender. A straightforward way to measure 
homophily in the mutiplex data is to compare the numbers of directed links between all gender-combinations 
in all networks (MM, MF, FM, FF) to the corresponding numbers surrogate networks where the gender of 
nodes is randomized. Table~\ref{tab:weights}, lower panel, shows the result of this comparison. Positive and 
negative Z-scores (see SI) denote the number of standard deviations by which a link type is over- or 
under-represented, with respect to the randomized case.
To measure statistical significance of differences between different combinations of genders, we compare 
each real network to 1,000 surrogate networks where the gender of the nodes was randomly reshuffled, 
leaving the topology of the network intact.
Clearly, female-to-female trading  and communication are the most significantly overrepresented link types, 
with a Z-score of approximately 4 sigmas.  Male-to-female trades (Z=2.7) and communication (Z=2.7) are also strongly 
over-represented, whereas  the opposite, female-to-male trades and PMs is much less substantial (Z=1.4 and 
1.6, respectively). Male-to-male trades and communication are highly under-represented (Z= -2.3 for both cases). 
Negative link types show no  significant tendency in either direction.

\begin{figure}[!t]
    \begin{center}
        \includegraphics{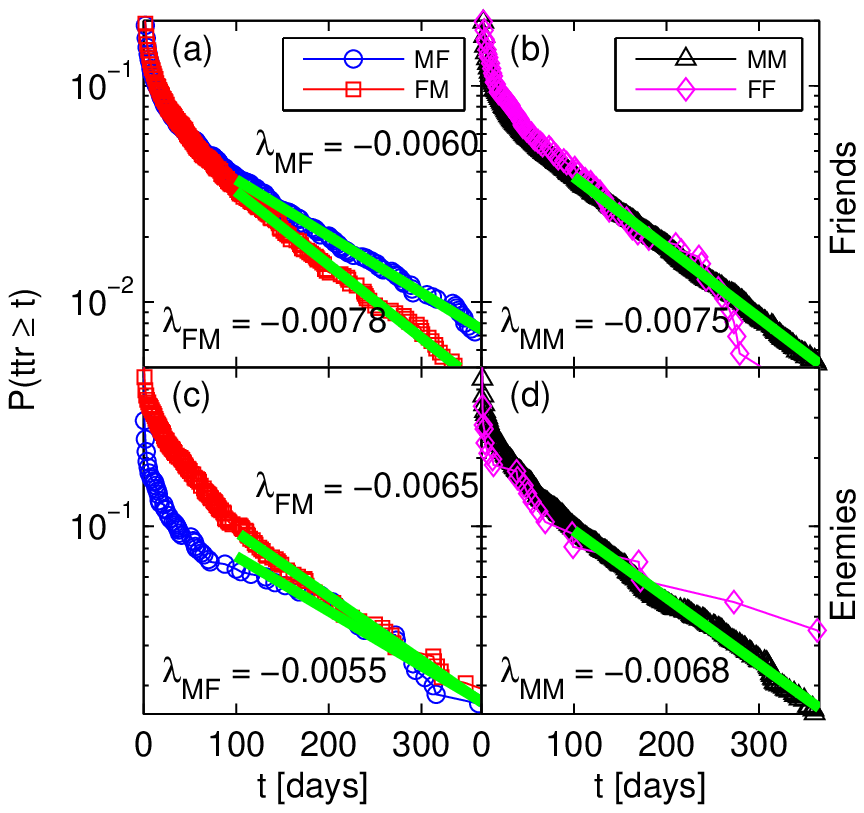}
    \end{center}
    \caption{{\bf Distributions of time-to-respond for different genders.} (a)  Cumulative probability distributions for the time-to-respond (ttr) for females to 
    	reciprocate a friendship link, given the initiator was male (MF), and vice versa (FM). 
	Probabilities fall sub-exponentially in the first 30 days, later  exponentially, $P(\mathrm{ttr}\geq\!t) \sim e^{-\lambda t}$ with long-time 
	decay rates $\lambda$ depending on gender pairs: 
	Males are much faster to reciprocate female friendship initiatives ($\lambda_{\mathrm{FM}} = -0.0078$) than the other way round ($\lambda_{\mathrm{MF}} = 	-0.0060$). (b) Situation for equal sex reciprocation MM, and FF. Here reciprocation decay times are  in between the MF and FM case.
	(c) Cumulative ttr distributions for enemy links. For  enemy markings, males are considerably slower to reciprocate within the first 180 days 
	if the initiator was a female  than vice versa. 
	(d) Situation for equal sex reciprocation of enemy links. Fit ranges are 100 to 365 days, fits of FF curves were not feasible.
\label{fig:timetoreciprocate} }
\end{figure}

{\bf Males respond fast (slow) to female friendship (enmity) initiatives.}
We measure the time (days) it takes for individuals to reciprocate an action of a given type. For all pairs $(i,j)$ where 
an action was reciprocated, the time-to-reciprocate (ttr) is the number of days from initial marking to reciprocation. 
In Fig.~\ref{fig:timetoreciprocate} (a) and (b) we show the cumulative distribution functions for the time-to-reciprocate 
for friendship and enmity links, for the four possible gender permutations:  MM, FF, MF, FM. Here the first letter denotes 
the gender of the individual who placed the initiating link, the second letter denotes the gender of the one who reciprocated. 
Reciprocation time probabilities follow a sub-exponential decrease for about the first 30 days after a link was initiated, 
beyond which distributions become approximately exponential, $P(\mathrm{ttr}\geq\!t) \sim e^{-\lambda t}$. 
The decay rate $\lambda$ for friendship reciprocation with  male initiation followed by female reciprocation is 
$\lambda_{\mathrm{MF}} \approx -0.0060$. For female initiation and male reciprocation it is $\lambda_{\mathrm{FM}} 
\approx -0.0078$. The MM rate lies between, close to the FM case, $\lambda_{\mathrm{MM}} \approx -0.0075$, (b). 
Correspondingly, the half-life for MF reciprocation is about 116, for FM it is only 89 days. The situation changes for the 
enemy links. Here a big difference is found for the short times: within the first 150 days of an enemy marking males 
reciprocate much less than females. The approximate exponential tails have a $\lambda_{\mathrm{FM}} \approx -0.0065$ 
and  $\lambda_{\mathrm{MF}} \approx -0.0055$, respectively. The ttr-distributions for the other actions are found in the SI.

\begin{figure}[!t]
    \begin{center}
        \includegraphics{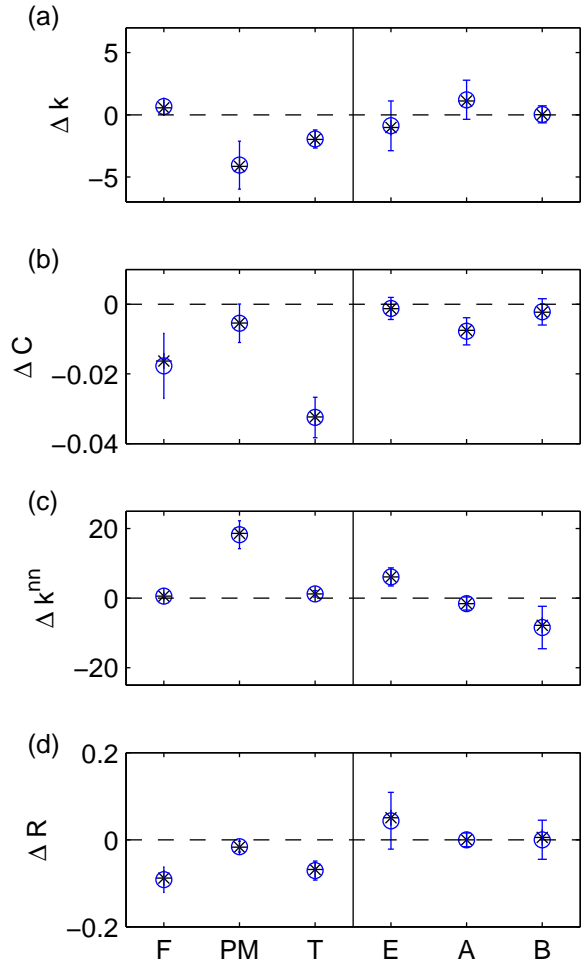}
    \end{center}
    \caption{{\bf Differences of network properties between male and female players ($M-F$) on the last day 856.} (a) average degree $\bar{k}$, (b) clustering coefficient $C$, (c) average neighbour degree $\bar{k}^{nn}$. Females have higher average degrees in PM and trade networks, as well as a higher clustering coefficient in trades, but considerably lower $\bar{k}^{nn}$ for PMs. (d) reciprocity $R$ between male-male and female-female links on the last day 856. 
Females are substantially more reciprocal in friendships and trades. Errorbars denote the standard deviations of the network measures obtained from the 8 male control groups (each of the size of the female group), illustrating the substantial differences. A statistical t-test supports these results by significant rejection of the equal means hypothesis. For definitions see Methods. \label{fig:propertydifferences}}
\end{figure}

{\bf Gender Differences in Networking.} 
Figure \ref{fig:propertydifferences} shows gender differences in four network properties as observed on day  856, 
illustrating the clear evidence that males and females structure their local MPNs in different ways. We compute the 
average degree $\bar{k}$, the clustering coefficient $C$, and the average neighbour degree $k^{\mathrm{nn}}$ 
(see SI) independently for the 8 male control groups and the female group, see Methods. We then subtract the 
value for the female group from the average of the 8 control groups. This difference is indicated in 
Fig.~\ref{fig:propertydifferences} by the letter $\Delta$. 
Errorbars indicate the standard deviations of the means of the control groups. 
A statistical t-test supports these results by significant rejection of the equal-means hypothesis, see Methods and SI.

{\em Females have more communication partners.}
As seen in Fig.~\ref{fig:propertydifferences} (a) females have about 15\% more communication and trading partners 
(degree) than males.

{\em Females organize in clusters.}
Female trading networks show a clustering coefficient (see SI) that is about 25\% higher than the one of males,  
Fig.~\ref{fig:propertydifferences} (b). This means that females tend to trade with people who trade among themselves. 
Also the clustering of female friendship networks is significantly higher than those of males, showing a preference for 
stability \cite{granovetter1973swt} in local networks. Surprisingly, also in the case of attacks females are more likely to 
attack people who are in conflict with each other than  males. 

{\em Males prefer well connected communication partners.}
From Fig.~\ref{fig:propertydifferences} (c) we learn that the communication partners of males  have more communication 
partners than the communication partners of females (for the neighbour degree $k^{nn}$, see SI). The same tendency  
is seen for the male enmity networks, meaning that the typical enemy of a male player has more enemies than the typical 
enemy of a female. In relative terms, both effects are in the 10\% range.

{\em Females reciprocate friendships.}
Females invest more effort in reciprocating links. Fig.~\ref{fig:propertydifferences} (d) reports that females reciprocate 
20\% more friendship and trading links than their male counterparts. For communication and negative links there are 
no  substantial gender differences observable. For the measure $\Delta R$ we considered male-male and female-female 
links. More details are found in the SI.

\section{Discussion}

We quantitatively studied gender-specific behavior and network organization in a society of humans engaged in a 
virtual world of an online game. We find clear differences of how females and males structure and manage their local 
MPNs. Females show a significantly higher average degree $\bar{k}$ in their communication networks, however, the 
communication partners of females have a significantly lower average degree than those of males, i.e. females have 
more communication partners, while males tend to have better connected ones. The positive MPNs of females are far 
more clustered than those of males. The combination of  higher clustering coefficients with  lower average neighbour 
degrees $k^{nn}$  indicates that females manage their local MPNs to be tighter and more compact. On the behavioral 
level females are more engaged in the reciprocation of their positive relations, again signaling that they invest more in 
stable and secure networks. This is also reflected in their less risk-taking behavior and a better overall economic 
performance in the game.

So far only a limited body of data-driven works on gender-specific network organization exists. A particular exciting one is 
a study on the use of  classic communication media (telephone)  \cite{smoreda2000gud}, where it was found that women 
call more frequently and spend about twice as much time on the telephone than men. We find a comparable effect. Female 
players send about 25\% more messages (0.74 per day) than males (0.60 per day). In \cite{smoreda2000gud} a gender 
homophily effect  \cite{mcpherson2001bfh} was found for both sexes. We find a comparable effect for females only where 
female-to-female links in the communication network are strongly over-represented with respect to pure chance. In contrast 
to \cite{smoreda2000gud} we  find that links between pairs of males are less likely than expected by chance. This tendency 
might be caused by the male-dominated  game environment in which females may be treated far better than males, which is 
a known phenomenon \cite{griffiths2003bsc}, and therefore also receive more messages. Inter-gender response time 
distributions reveal that females respond much slower to male friendship initiatives, than males do to female ones. We find 
clear evidence that females `attract' positive behavior to a much higher degree than males. 
It is important to note that players have neither direct information about -- nor influence on -- the links of their neighbours 
(except for their own). Therefore, the observed differences in network patterns such as clustering coefficients and average 
neighbour degree, can not be explained through social group dynamics, such as e.g. in the friendship dynamics in 
{\em Facebook}, but instead constitute strong direct evidence for biological gender-specific  differences in networking behavior.

The direct implications of the presented work to real societies has to be discussed with some care. First, it is not a priory 
obvious how representative  MMOG societies are for real human societies, and second, how much does gender swapping 
influence the results? The only way to determine to what extent MMOGs are good models for `real' society is the direct 
comparison of the MPNs in games with those observable in the `real' world. Previous results on the same Pardus dataset 
show amazingly good agreements  \cite{szell2010msd} between statistical properties of the communication network of 
players and mobile phone networks \cite{onnela2007sts, lambiotte2008gdm}. Also in \cite{szell2010msd} the evolution of 
network measures and of network growth patterns, including network densification, was shown to strongly overlap with data 
from `real-world' societies \cite{leskovec2007ged,leskovec2008mes}. In further work \cite{szell2010mol} additional equivalent 
results have been found in the study of sociological hypotheses on structural balance \cite{leskovec2010sns}.
We have not explicitly conducted a survey on the extent of gender swapping in Pardus, however we have reason to believe that 
we are in the same ballpark as comparable MMOGs \cite{grosman2010esg}, meaning that we expect about 15 \% of players 
whose avatar has a different gender than their biological player. For all of our conclusions this fraction can be regarded as marginal. 

In conclusion, the substantial gender effects in different types of networking and reciprocation behavior that we find in the 
online world of Pardus, suggest two main implications. First, if we assume our findings to hold in online environments 
in general and if the majority of players choose their online gender in accordance  with their biological sex, this implies 
that we can find and explore traces of biological human behavior in online environments. As previous 
studies on mobility \cite{szell2012ums} or social organization hypotheses \cite{szell2010msd, szell2010mol} have shown, 
the type of medium and actual physical contact are of limited relevance for a wide range of social behavior.
Our results imply an impact on the understanding of relation-formation between individuals of different and the same genders, 
and thus eventually might help to understand reproductive strategies behind the behavior of finding and competing for mating partners 
\cite{palchykov2012sdi}. Second, our setup demonstrates that online media increasingly blur gender roles and can render 
the biological sex of social actors less important. Without tangible signals, \emph{representing} the behavior of a male or 
female may be more important here than actually having male or female sex. Lacking information on sex does not enable 
us to disentangle sex--gender effects: Does a male acting as a female behave differently from a female who chooses a 
female online persona? We do not know -- further data and studies on gender-specific offline--online behavior are needed 
to answer these specific details. Nevertheless, our results could have immediate applications both offline and online, in 
particular in control of gender-specific information spreading, in marketing, or for smooth implementation of online social 
networks and dating sites \cite{bingol2012amw}.

\section{Methods}
\subsection{The Social Multiplex Network Data}
For the MPN we considered all players of the {\em Artemis universe} who played within the time interval from 
day 1 to day 856 and who performed at least one communication event, a trade, an attack, or a bounty request, 
or who were involved in at least one friendship or enmity relation on day 856 (the trade relation studied here is 
all ship-to-ship trades, reflecting the business relations in which trust is important -- this is slightly different to 
trade networks constructed from ship-to-building trades which can be of more impersonal nature 
\cite{szell2010msd, szell2010mol}). We discard all players who existed for less than 1 day. The communication-, 
trade-, attack-, and bounty networks were accumulated over time, enmity and friend networks were taken as snapshots 
of day 856. This leads to 23,872 players (nodes) in the MPN with 21,264  males and 2,608 females. For the 
time-to-respond calculations we used also all players involved in reciprocated friendship and enmity markings 
over the 856 days, increasing the total number of players to 34,210 (30,607 males and 3,603 females). For the 
calculations of player achievements, we used those 6,548 players who played on day 856. 

\subsection{Control Groups}
At any time, there are roughly 8 to 9 times more male than female players.
For highlighting substantial gender differences, we randomly divided the set of male players into 8 non-overlapping control groups, 
each of the size of the female group (3,603). We compute mean and standard deviation of these 8 control samples for comparing 
network- and performance measures in Fig.~\ref{fig:propertydifferences} and in Supplementary Table~I. Following the central limit 
theorem, with a large enough sample size the standard deviation of the sample means follows a normal distribution regardless of the 
shape of the parent population. Since the number of samples here is very low (8), the deviations of the sample means are to be 
treated with care. For rigorous statistical testing however, we use a two-sample t-test, see below.

\subsection{Statistical Technique for Hypothesis Testing}
For directly comparing values between males and females, we use a standard two-sample t-test for testing 
the null hypothesis $H_0$ that two distributions have equal means, against the alternative hypothesis $H_1$ 
that the means are not equal. We assume equal variances and use two tails, i.e.~no preference of testing for 
one specific gender to have larger values than the other. Thus the t-test computes a pooled standard deviation
\begin{equation}
s = \sqrt{\frac{(n_m-1) \mathrm{var}_m + (n_f-1) \mathrm{var}_f}{n_m + n_f - 2}}
\end{equation}
where $n_m$, $n_f$ are the sizes of male and female values considered, and $\mathrm{var}_m$, $\mathrm{var}_f$ 
their variances, respectively. The test statistic is then given by
\begin{equation}
t = \frac{\bar{m}-\bar{f}}{\sqrt{\mathrm{var}_m/n_m + \mathrm{var}_f/n_f}}
\end{equation}
where $\bar{m}$ and $\bar{f}$ denote the means of the male and female values.

\begin{acknowledgments}
\subsection{ACKNOWLEDGEMENTS}
The authors acknowledge support from the Austrian Science Fund FWF P23378. We thank Benedikt Fuchs for assistance in data analysis, and Bernat Corominas-Murtra for stimulating discussions.
\end{acknowledgments}


\end{document}